\documentclass[reprint, 
amsmath,
amssymb, 
floatfix, 
aps,
prb,
superscriptaddress]{revtex4-2} 
\usepackage{graphicx}

\usepackage[utf8]{inputenc}
\usepackage{xcolor}
\definecolor{codegreen}{rgb}{0,0.6,0}
\definecolor{codegray}{rgb}{0.5,0.5,0.5}
\definecolor{codepurple}{rgb}{0.58,0,0.82}
\definecolor{backcolour}{rgb}{0.95,0.95,0.92}
\definecolor{urlblue}{HTML}{007bff}

\usepackage{bm}
\usepackage{bbm}
\usepackage{orcidlink}
\usepackage{microtype}
\usepackage{enumitem}
\usepackage{braket}
\usepackage{mathtools}
\usepackage{wasysym}
\usepackage{times}
\usepackage{appendix}
\usepackage[capitalise]{cleveref}
\input{macros.sty}
\usepackage{orcidlink} 
\usepackage[normalem]{ulem}

\hypersetup{
    colorlinks = true,
    linkcolor  = urlblue,
    citecolor  = urlblue,
    urlcolor   = urlblue,
}

\usepackage[mathlines]{lineno}

\begin{document}


\title{Altermagnetism Without Crystal Symmetry}

\author{Peru d'Ornellas\orcidlink{0000-0002-2349-0044}}
\thanks{These authors contributed equally.}
\affiliation{\small Universit\'e Grenoble Alpes, CNRS, Grenoble INP, Institut N\'eel, 38000 Grenoble, France}
\author{Valentin Leeb\orcidlink{0000-0002-7099-0682}}
\thanks{These authors contributed equally.}
\affiliation{Department of Physics TQM, Technische Universit{\"a}t M{\"u}nchen, James-Franck-Stra{\ss}e 1, D-85748 Garching, Germany}
\affiliation{Munich Center for Quantum Science and Technology (MCQST), 80799 Munich, Germany}
\author{Adolfo G. Grushin\orcidlink{0000-0001-7678-7100}}
\affiliation{\small Universit\'e Grenoble Alpes, CNRS, Grenoble INP, Institut N\'eel, 38000 Grenoble, France}
\author{Johannes Knolle\orcidlink{0000-0002-0956-2419}}
\affiliation{Department of Physics TQM, Technische Universit{\"a}t M{\"u}nchen, James-Franck-Stra{\ss}e 1, D-85748 Garching, Germany}
\affiliation{Munich Center for Quantum Science and Technology (MCQST), 80799 Munich, Germany}
\affiliation{\small Blackett Laboratory, Imperial College London, London SW7 2AZ, United Kingdom}

\date{\today}

\begin{abstract}
    Altermagnetism is a collinear magnetic order in which opposite spin species are exchanged under a real-space rotation. Hence, the search for physical realizations has focussed on crystalline solids with specific rotational symmetry. Here, we show that altermagnetism can also emerge in non-crystalline systems, such as amorphous solids, despite the lack of global rotational symmetries. We construct a  minimal Hamiltonian with two directional orbitals per site on an amorphous lattice with interactions that are invariant under spin rotation. Altermagnetism then arises due to spontaneous symmetry breaking in the spin and orbital degrees of freedom around each atom, displaying a common point group symmetry. This form of altermagnetism exhibits anisotropic spin transport and spin spectral functions, both experimentally measurable. Our mechanism generalizes to any lattice and any altermagnetic order, opening the search for altermagnetic phenomena to non-crystalline systems.
\end{abstract}

\maketitle

{\it Introduction---}
Symmetry allows us to organize nature's staggering variety of magnetic materials into classes with different macroscopic properties. Ferromagnetic (FM) order breaks time reversal symmetry (TRS), leading to electronic structures with uniformally spin-split bands~\cite{landau_lifshitz_ferro_antiferro}. Antiferromagnetic (AFM) order preserves a combination of TRS and lattice translation (or inversion), and thus has a spin-degenerate band structure~\cite{neel_magnetism_1971}. 
The effort to find a complete symmetry classification in terms of real-space and spin transformations~\cite{corticelli2022spin,liu2022spin,yang2024symmetry}  has led to the proposal of a new collinear order, dubbed altermagnetism (AM)
~\cite{smejkal_beyond_2022, smejkal_emerging_2022}. Like AFM, altermagnets, historically known as $d$-wave magnets or higher orders thereof \cite{pomeranchuk_stability_1958,wu_fermi_2007,ahn2019antiferromagnetism}, have zero net magnetization. However, unlike AFM, their electronic band-structures are spin-split. Altermagnetic order preserves the combination of spin-flip and real-space rotational symmetry while breaking inversion and lattice translation symmetries between the spin sublattices \cite{smejkal_emerging_2022}.

The rotation symmetry classifying AMs has an ambiguity in its definition. We can conceive of two distinct definitions of a real-space rotation operator: a rotation of the spatial coordinates of the entire system and a local rotation that acts individually on the orbital degrees of freedom around each site. 
Studies on AM have predominantly focused on the former definition, where a \textit{crystallographic} rotation symmetry is enforced by sites on different sublattices having local environments related by a rotation \cite{smejkal_emerging_2022,mazin_prediction_2021,smejkal_crystal_2020}, for example on an undistorted bipartite lattices with a unit cell with two AFM sublattices that are related neither by inversion nor translations but by rotations~\cite{fischer2011mean,brekke2023two,bose_altermagnetism_2024,durrnagel2024altermagnetic,das_realising_2024,kaushal2024altermagnetism,li2024d}.

However, recent work has found that AM can be quantified using a multipolar order parameter \cite{bhowal_ferroically_2024,mcclarty_landau_2024}, which can encode the \textit{orbital} symmetry of a magnetic configuration around a single site. Furthermore, rather than relying on an explicit breaking of spin and real-space rotational symmetry, it is possible for AM ordering to emerge from spontaneous symmetry breaking in an interacting system without crystallographic sublattice anisotropy~\cite{leeb_spontaneous_2024}. Here, real-space symmetry breaking arises due to the anisotropic shape of electronic orbitals themselves, rather than any symmetry of the lattice structure.

This motivates the question: Is crystalline lattice symmetry necessary---or is it possible to realise AM only through combined spin and orbital ordering on a lattice that shares none of the relevant symmetries?
A number of crystalline electronic phases have recently been extended to amorphous lattice geometries, such as topological insulators~\cite{mansha_robust_2017,xiao_photonic_2017,agarwala_topological_2017,Poyhonen2017,mitchell_amorphous_2018,grushin_topological_2023,  marsal_topological_2020, Mukati2020,corbae_amorphous_2023, corbae_observation_2023,kim2023fractionalization} and spin liquids \cite{grushin2023amorphous, cassella_exact_2023}. It is unclear if it is possible to construct an altermagnetic ground state in an arbitrary amorphous lattice geometry, or whether---as with AFM---amorphous structure generally leads to glassy physics \cite{zallen_physics_2008,coey_amorphous_1978}.

In this work, we construct a minimal toy model for atomic altermagnetism, i.e., the multipolar moment of magnetization originates from a single site or atom, which emerges through spontaneous symmetry breaking in a system that preserves rotational symmetry.
We construct a Hamiltonian on an amorphous lattice that respects on-site rotational symmetry in both spin and orbital space.
We calculate the full phase diagram using real-space mean field theory and quantify two observables that act as a signature of altermagnetic ordering, the spin-resolved spectral function and anisotropic spin conductance. 
Our mechanism is sufficiently general to be applicable to any non-crystalline lattice. Since it is created {\it ferroically}, with every site in the lattice placed in the same configuration in spin and orbit, it does not suffer from geometrical frustration. Furthermore, this mechanism can lead to any symmetry of the altermagnetic order-parameter.

{\it The Model---}%
\begin{figure}[t]
    \centering
    \includegraphics[]{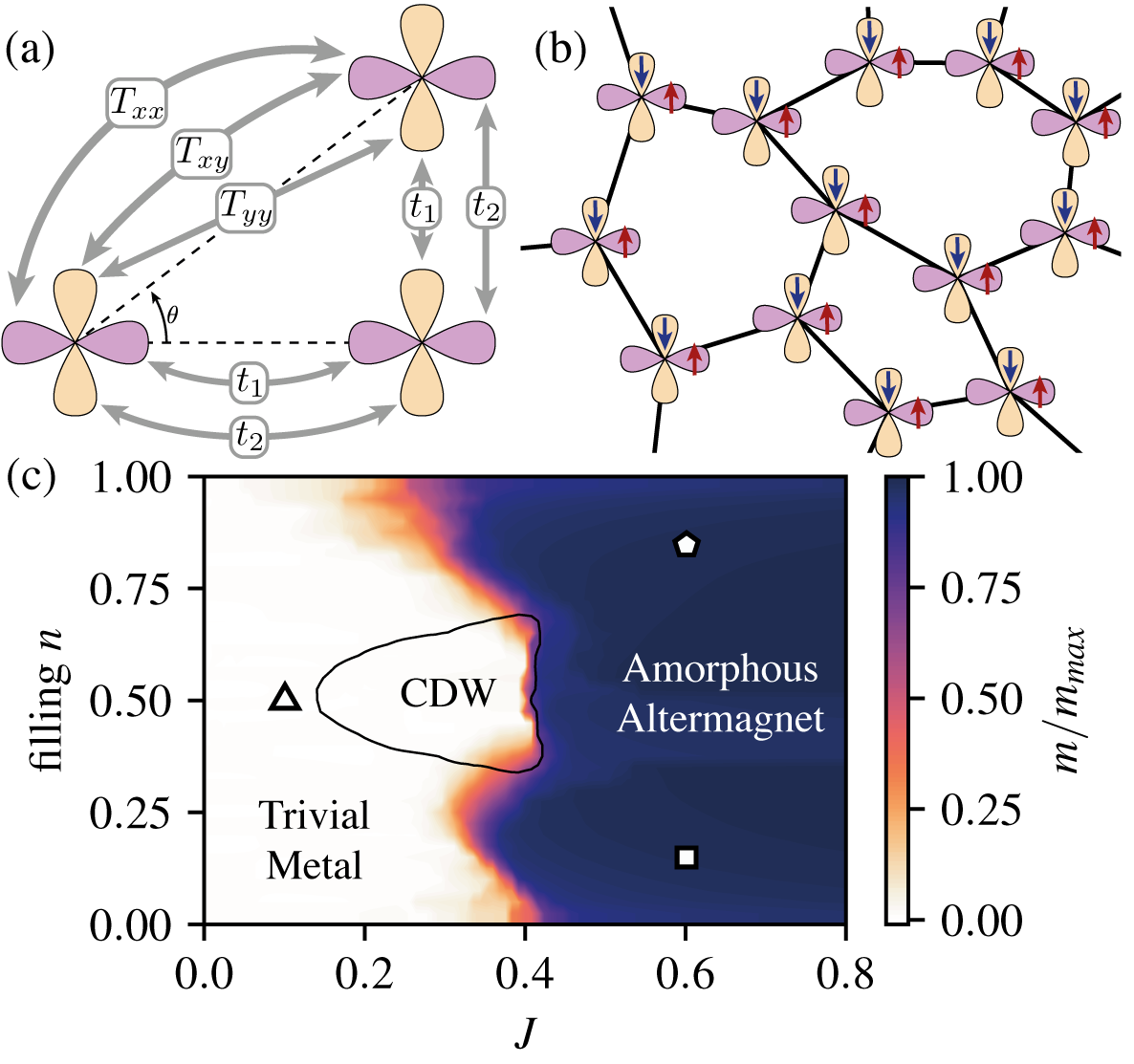}
    \caption{
    (a) 
    Adjacent orbitals are coupled with an orbital-dependent hopping parametrised by two terms. 
    A strong $t_1$ acts along the orbital's axis ($\sigma$-bonding) and a weak $t_2$ acts perpendicular ($\pi$-bonding). At intermediate angles the hopping interpolates between the types of bonding via a matrix $T(\theta)$. 
    (b) 
    A section of an amorphous lattice with an altermagnetic state at half filling. Each site hosts two orbitals, occupied by spin-up and spin-down electrons, respectively. The state is invariant under a combined $\mathcal T$ spin flip and \textit{local} $C_4$ rotation of the orbitals around each site. 
    (c)
    A phase diagram for a three-coordinated amorphous lattice in filling $n$ and the interaction coupling strength $J$. The filling is parametrised such that $n=1$ is a fully filled system, with four fermions per lattice site. Three phases are shown: a trivial metal, 
    an amorphous altermagnet, and a charge-density wave.
    The colouring represents the mean alter-magnetization $m$, normalized to its maximal value $m_{\textit{max}}=2-|4n-2|$. Labelled points are discussed in the text.
    }
    \label{fig:1}
\end{figure}%
We consider a $t$-$J$--like model of interacting fermions on an amorphous three-coordinated lattice. Each site features two orbitals, labelled $x$ and $y$. 
We choose $d_{xz}$ and $d_{yz}$ orbitals for concreteness, see \cref{fig:1}a. However, the exact shape of the orbitals does not matter, provided that they break rotation symmetry and are mapped onto one another under a $C_4$ rotation. 
The Hamiltonian $H=H_K+H_{\textit{Int}}$ comprises two parts, a kinetic term $H_K$ which allows hopping between sites, and a nearest-neighbour spin orbital interaction term $H_{\textit{Int}}$ specified below.

As a proof of principle, we choose couplings which realize a compensated order with spin-split band structure. The kinetic term is a nearest-neighbor hopping in which different orbitals have a preference for hopping in either the $x$ or $y$-direction, which is characteristic for anisotropic $d$-orbitals. This ensures that the system has an orbital-split band structure. Next, we wish to introduce a pairing between orbital states and spin states, such that the orbital-split band structure is converted into a spin-split band structure. This is done by an interaction that prefers to have alike spins  in the same orbitals, and orthogonal spin species in orthogonal orbitals. Thus, one spin species (for example spin up) preferentially populates $x$ orbitals, and so hops preferentially in the $x$ direction, with the opposite for spin down. Finally, we aim to construct an interaction that respects both spin and rotation symmetry, such that the altermagnetic state emerges as a consequence of spontaneous symmetry breaking. 

The kinetic part of the Hamiltonian represents an orbital-dependent, spin-degenerate nearest neighbour hopping, which we express in the basis $\Psi_j^\dag = (c^{\dag}_{jx\uparrow}, c^{\dag}_{jy\uparrow}, c^{\dag}_{jx\downarrow},c^{\dag}_{jy\downarrow})$ as
\begin{align}
    H_{K} 
    &= \sum_{\braket{jk}}
    \Psi_{j }^\dagger
        [ T(\theta_{jk}) \otimes \1]
    \Psi_{k}^{},
\end{align}
where $\theta_{jk}$ is the angle a given bond makes with the $x$ axis. The matrix $T$ has two parameters, $t_1$ and $t_2$, which encode the bond-direction dependence of the overlap between $x$ and $y$ orbitals. For $\theta=0$ the $x$-orbitals have a larger overlap $t_1$, whereas the $y$-orbitals have a smaller overlap with $t_2<t_1$. For $\theta=\pi/2$ these hoppings are reversed. In between, the hopping, described by the matrix $T(\theta)$, can be determined by rotating the orbitals to a basis aligned with the bond, coupling with $t_1$ and $t_2$, and then rotating back to the original frame. This is shown in \cref{fig:1}a, with an explicit formulation given in the End Matter. 

Here, the difference between $t_1$ and $t_2$ encodes the degree to which $x$ orbitals preferentially hop in the $x$ direction, and $y$ orbitals preferentially hop in the $y$ direction. As long as $t_1 \neq t_2$, this hopping ensures an orbital-split band structure. In our model all hoppings have the same strength, regardless of bond length. Of course, using a distance-dependent hopping is microscopically more accurate, however this does not change the qualitative physics. Furthermore, in many amorphous materials the distance, and coupling, between adjacent atoms is determined by the material's chemistry. Thus, they are often close to uniform, and the effects of disorder arise in the connectivity of the lattice \cite{zallen_physics_2008,weaire_electronic_1971}. Thus, in the interest of keeping the minimal model as simple as possible, we do not include it.

The interacting part is a ferromagnetic Heisenberg-like combined spin and orbital interaction which respects global $SU(2)\times SU(2)$ spin-orbital symmetry
\begin{align} \label{eqn:interaction}
    H_{\textit{Int}}
    &= -J\sum_{\substack{\braket{jk} \\ \alpha \beta}}
        \left(\Psi_{j}^\dag 
        \tau^\alpha
        \otimes 
        \sigma^\beta 
        \Psi_{j} \right)  
        \left(\Psi_{k}^\dag 
        \tau^\alpha
        \otimes 
        \sigma^\beta 
        \Psi_{k} \right)
        -
    n_{j}
    n_{k},
\end{align}
where $n_j = c^\dagger_{j\alpha s}c_{j\alpha s}$ is the total on-site occupation operator and $\sigma^\alpha$ ($\tau^\alpha$) are Pauli matrices acting on the spin (orbital) subspace. Note that in the single fermion-per-site limit \cref{eqn:interaction} is analogous to a \textit{ferromagnetic} Kugel--Kohmskii interaction (normally the Kugel--Kohmskii interaction is antiferromagnetic), with the form $(\bm S_j \cdot \bm S_k)(\bm \tau_j \cdot \bm \tau_k)$ \cite{kugel_jahn_2983}. For $J>0$, this interaction acts as a ferromagnetic Heisenberg coupling between aligned orbitals, and an antiferromagnetic coupling between orthogonal orbitals. Thus, the interaction energy is minimised in a mean-field sense by placing the same spin species in all $x$-orbitals, and the opposite spin species in all $y$-orbitals.

\begin{figure*}[ht]
    \centering
    \includegraphics[width=\textwidth]{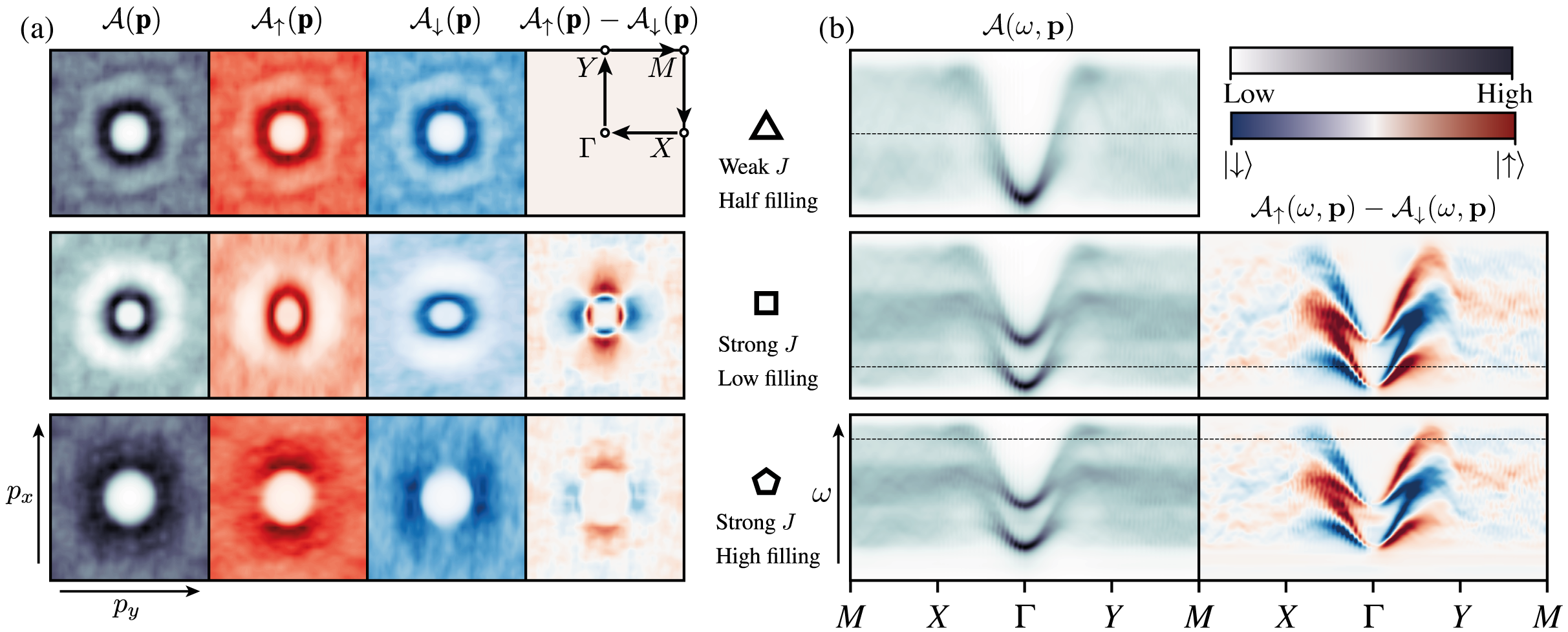}
    \caption{ Spectral functions for three example points in the phase diagram, labelled by three polygons that correspond to the points in \cref{fig:1}c. The first row ($\triangle$) shows the trivial metal phase, with weak $J/t$ at half filling. The second row ($\Square$) shows an altermagnetic phase with strong $J/t$ at low filling, where the states at the Fermi level are close to $\Gamma$. The third row ($\pentagon $) shows an altermagnetic phase with strong $J/t$ at high filling. Here the states close to the Fermi level are far from $\Gamma$ so are not well-approximated by plane waves. 
    (a) Spectral density as a function of $\textbf p$ at the Fermi level $\epsilon_F$ (shown as dashed line in panel b). Four cases are shown, the total density, the densities of spin-up and spin-down respectively, and the difference between spin-up and spin-down. The difference vanishes in the non-altermagnetic case. 
    (b) Spectral density as a function of energy and momentum on a path around the quasi-Brillouin zone (shown in the top right panel of subfig.~a. Two cases are shown, the overall spectral density and the spin-difference density. The quasi-Brillouin zone $[-\pi/\bar a,\pi/\bar a]^2$ and its high symmetry points are defined via the average lattice spacing $\bar a$, which is the relevant length scale in an amorphous solid~\cite{Ciocys_2024}.
    }
    \label{fig:2}
\end{figure*}%

{\it Symmetry Breaking---}%
This coupling respects global $SU(2) \times SU(2)$ symmetry in the combined spin and orbital space---that is we are free to rotate our basis in spin space or orbital space. However, the full Hamiltonian $H$ has only a global $SU(2) \times SO(2)$ symmetry, due to the kinetic part $H_K$. Each bond chooses a rotated combination of orbitals to couple preferentially, such that \textit{on average} the system is invariant under $SO(2)$ rotation in real space (or equivalently of the orbitals).

To satisfy the coupling, we choose a state in orbital space and a spin species, and populate all such orbitals with the chosen spin species. Next, we populate all orthogonal orbitals with the orthogonal spin species. The choice of states in both spin and orbit is arbitrarily picked by spontaneous symmetry breaking.  
We assume that both of these are in the basis of spin $\sigma^z$ and orbital $\tau^z$, hence we may discard all other operators. Thus, the interaction simplifies, and is written in terms of the total occupation $n_j$ and the alter-magnetisation $m_j := \Psi_j^\dag \sigma^z \tau^z\Psi_j = n_{jx\uparrow} -n_{jy\uparrow} -n_{jx\downarrow}+n_{jy\downarrow}$, arriving at the interaction,
\begin{align}\label{eqn:mean_field_interaction}
    H_{\textit{Int}} \rightarrow - J\sum_{\braket{jk}}\left (
     m_j
     m_k
    - 
     n_j
     n_k \right ).
\end{align}
When $m_j$ has a non-zero expectation value, the system separately breaks rotation symmetry, which exchanges orbitals, and TRS which exchanges spin species. However, the system still preserves their combination $C_4\mathcal T$, since this operator commutes with $m_j$ on all sites \cite{giuli_altermagnetism_2025}. A spontaneous symmetry breaking of spin and orbital is not unprecedented in non-crystalline solids. Similar symmetry breaking phenomena, albeit non-magnetic, have been observed in amorphous Bi$_2$Se$_3$~\cite{corbae_observation_2023,Ciocys_2024}, whose surface states display a measurable spin-orbit texture~\cite{corbae_observation_2023}. This winding is presumably chosen by non-universal surface phenomena, such as the surface potential, and we expect that similar effects could emerge in our context.

Note that this decoupling is most applicable in the limit of strong coupling $ V \gg t_1, t_2$. At weak coupling, we would expect that the hopping terms would compete with the interaction term in deciding the local orientation of symmetry breaking at each site. However, since the hopping terms preserve $SO(2)$ on average, we expect that this effect will be weak and that for most of the parameter space our assumptions are valid. An exploration of this intermediate coupling regime is beyond the scope of this work. Finally, note that even in the disordered limit any finite alter-magnetisation would induce the features we discuss below. 

{\it Mean Field Decoupling---}%
We are now in a position to perform a mean field decoupling of the Hamiltonian, introducing expectation values for the number density $\braket{n_j}$ and alter-magnetisation $\braket{m_j}$. The decoupled interaction takes the following form,
\begin{align}\begin{aligned}
    H_{\textit{Int}}({\bf m}, {\bf n}) = 
    -J\sum_{\braket{jk}} 
    \braket{m_j} m_k 
    -\braket{n_j} n_k
    + (j\leftrightarrow k),
\end{aligned}
\end{align}
where we have neglected to include a constant energy shift of the form $[-\braket{m_j} \braket{m_k}+\braket{n_j} \braket{n_k}]$, since it has no effect on the ground state. We determine self-consistent values for $ \braket{m_j}$ and $ \braket{n_j}$ using an iterative real-space Hartree--Fock mean field method. This is done on lattices with $\mathcal{N}=400$ sites and periodic boundary conditions, performing up to 400 iteration steps to optimize the $2\mathcal{N}$ mean fields. 

{\it Phase Diagram---}%
We study the phase diagram, i.e. the average magnetization $m=\sum_j  \braket{m_j} /\mathcal{N}$, as a function of two parameters, the total filling $n$ and the interaction strength, $J$, setting $t_1 = 1$ and $t_2 = 1/2$. In each case we determine the ground state configuration of our mean fields and calculate the energy spectrum. Three phases are found, with the full phase diagram shown in \cref{fig:1}c. At low $J$, the system forms a trivial metal, where the four spin-orbital states are completely degenerate and the local alter-magnetisations $ \braket{m_j}$ are zero everywhere. 

In the center of the phase diagram the ground state forms a gapped charge density wave (CDW). Here, $m=0$ and so the interaction, \cref{eqn:mean_field_interaction}, acts as an effective nearest-neighbour repulsion, encouraging electrons to cluster on every other site. Since this is on an amorphous lattice, the lack of bipartiteness means that the CDW is geometrically frustrated, leading to defects in the ordering. However, the phase has no altermagnetic splitting, and we will not focus on it here. 

For large $J$ the system converges to non-zero alter-magnetization $m$. This breaks the degeneracy of the four spin-orbital states such that, for $m>0$, states $\ket{\uparrow x}$ and $\ket{\downarrow y}$ have lower energy, whereas $\ket{\uparrow y}$ and $\ket{\downarrow x}$ have higher energy. Thus the spin and orbital are locked to each other, i.e. $x$-orbital electrons carry spin-up and vice versa. Thus, the anisotropic transport of the two orbital species induces an anisotropic dispersion in spin, where spin-up electrons have higher mobility in the $x$-direction and spin-down in the $y$ direction.

The phase diagram is not symmetric in filling, because particle-hole symmetry is broken on the amorphous lattice which is not bipartite. At low filling, the states at the Fermi level are close to long-wavelength plane waves, which are effectively indifferent to the lattice geometry. On the other hand, at high filling the eigenstates at the Fermi level are extremely sensitive to lattice geometry. Going from low filling to high filling, we find a decrease of the critical interaction strength $J$, although around half-filling the CDW suppresses the altermagnetic phase. Interestingly, we observe that at high filling, where the disorder of the amorphous lattice is most relevant, the altermagnetic phase is most stable, suggesting that amorphous order serves to enhance the stability of the AM phase.

{\it Spectral Function---}%
The spectral function directly captures the spin-split bands of altermagnets and is a well-known probe for altermagnetism \cite{krempasky2024altermagnetic,lee2024broken,lanzini2025theory}.
Although crystal momentum is not a good quantum number on an amorphous lattice, we may still determine the momentum-space properties of the system by considering the overlap of the eigenstates with a set of plane waves, $\ket{\textbf p {s\mu}} = \mathcal N \sum_{\textbf x_j}e^{i \textbf p\cdot \textbf x_j}\ket{\textbf x_j {s\mu}} $, where $s$ denotes spin and $\mu$ orbital. These states do not form an orthonormal basis in an amorphous system, however they are almost orthonormal for momenta $\textbf p$ close to the $\Gamma$ point. Thus, we calculate the spin-resolved spectral function, which determines the probability of finding eigenstates $\ket {\lambda}$ with energy $\varepsilon_\lambda$ which have energy close to $\omega$ and overlap strongly with plane waves at a given $\textbf p$, 
\begin{align}
    \mathcal A_{s}(\omega,\textbf p) = -\frac{\eta}{\pi}\sum_{\lambda \mu}\frac{
        |\! \braket{\lambda | \textbf p {s\mu}}\!|^2
    }{
        (\omega - \varepsilon_\lambda)^2 + \eta^2
    },
\end{align}
where $\eta$ is a small spectral broadening term. We define $\mathcal{A}(\textbf p) = \mathcal{A}(\epsilon_F,\textbf p)$ to be the spectral function at the Fermi level $\epsilon_F$.
The spectral function $ \mathcal A_{s}(\omega,\textbf p)$ is measurable in spin-resolved angle-resolved photoemission spectroscopy (ARPES) experiments, even in non-crystalline solids and liquids~\cite{kim2011,corbae_observation_2023,Ciocys_2024}. This technique has revealed coherent electronic properties~\cite{corbae_observation_2023,Ciocys_2024} and spin-orbital textures of amorphous solids~\cite{corbae_observation_2023}.
In \cref{fig:2} we show the spectral function in momentum and energy for three points in the phase diagram, labelled with $\triangle$, $\Square$ and $\pentagon$ in \cref{fig:1}c.

The first case ($\triangle$) is that of a trivial metal, at half filling and weak $J=0.1$ . Here, all four spin-orbital states are degenerate. Thus, the spin up and spin down spectral functions are identical, and the spin-difference spectral function $\mathcal A_{\uparrow} - \mathcal A_{\downarrow}$ vanishes. Within the band, low energy states have a strong spectral weight close to the $\Gamma$ point. These states are effectively indifferent to the lattice geometry, since they are close to plane waves with wavelength much larger than the average lattice spacing. Thus, they may be understood as qualitatively similar to the low energy states found close to $\Gamma$ in a crystalline material. However, at large momentum, eigenstates are highly sensitive to the effects of lattice geometry and the plane waves provide no longer a good description. Thus, we expect that low filling configurations will be qualitatively similar to the crystalline case, whereas high filling configurations will not.

Next, we consider the altermagnetic phase, focusing on $J = 0.6$. We look both at low filling, $n = 0.1$ ($\Square$), and high filling $n = 0.8$ ($\pentagon$). At low filling we see similar results to those found in crystalline altermagnets: The spectral functions at the Fermi energy are Fermi-surface-like, forming a line where the band of low energy, long wavelength eigenstates crosses the Fermi energy. As expected in an amorphous system, the total spectral function is completely isotropic at any filling with no directionality chosen by the lattice structure. Looking at the spin-resolved spectral functions, we see a similar profile, however, with the substantial difference that a notion of direction survives, due to the spin-orbital ordering: The spin-up and spin-down Fermi surfaces are distorted into ovals with opposite orientations, leading to a spin-difference spectral function with $\mathcal T C_4$ symmetry. Looking at the energy-resolved spectral function, \cref{fig:2}b, we see that the non-zero $m$ has split each band into two, with the Fermi level only intersecting with the lower band, which consists predominantly of $\ket{\uparrow x}$ and $\ket{\downarrow y}$ states.

At high filling ($\pentagon$)  a similar phase is realized, with two split bands split by the non-zero $m$, and $A_\uparrow(\omega,\mathbf p)-A_\downarrow(\omega,\mathbf p)$ having a $C_4\mathcal T$ symmetry. However, the spectral function at the Fermi level is no longer Fermi-surface-like. The spectral weight is distributed over all momenta with a tendency of spin-up spectral weight to be located where $|p_y|>|p_x|$ and spin-down spectral weight to be located where $|p_x|>|p_y|$. The spin-differentiated spectral function captures this by broad sign changing features, which respect the for altermagnets typical $\mathcal T C_4$ symmetry. 

\begin{figure}[t]
    \centering
    \includegraphics[]{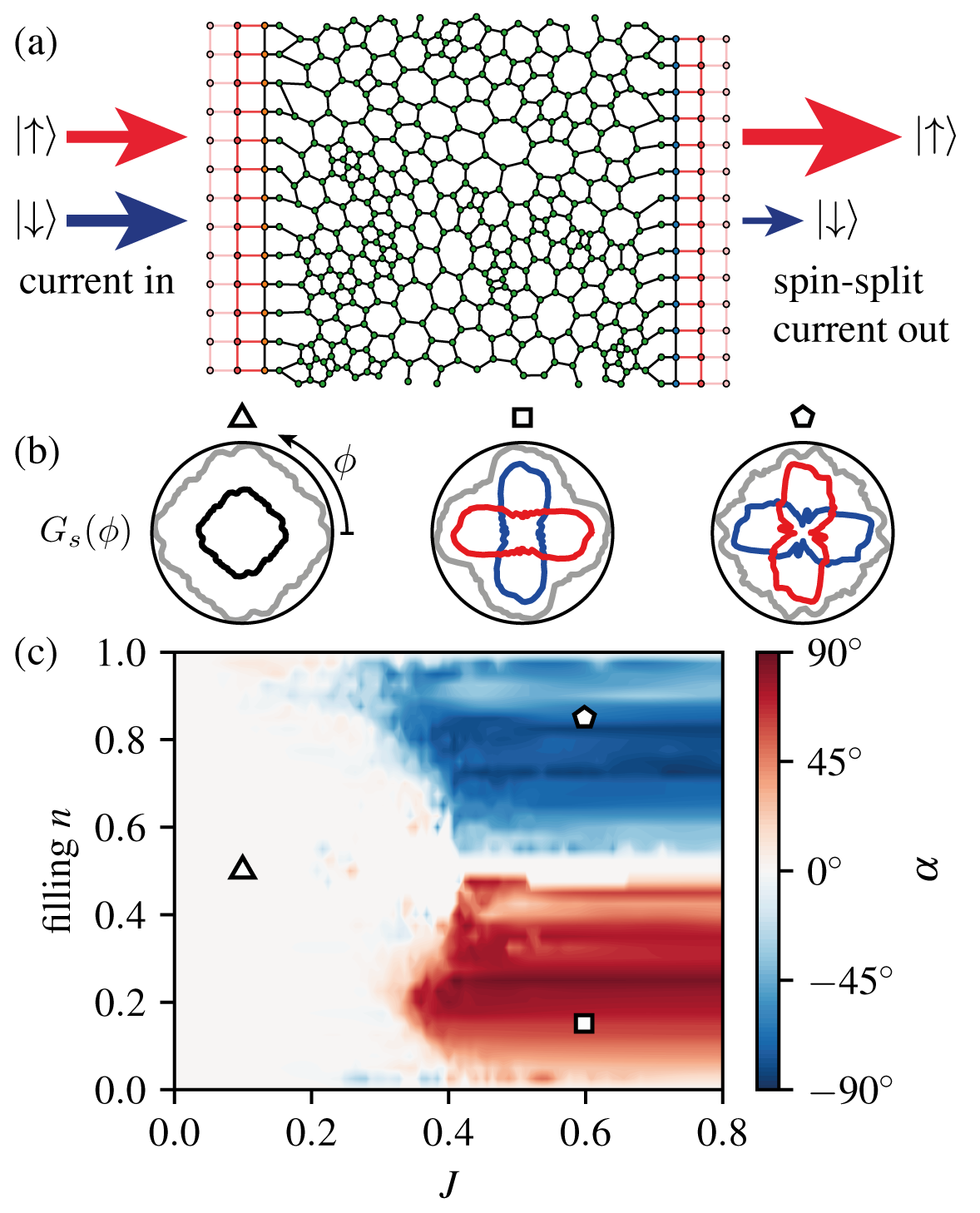}
    \caption{
    Spin conductance of an amorphous altermagnet. 
    (a) The set-up to evaluate the spin conductance consists of the mean-field-converged amorphous Hamiltonian, onto which we attach metallic leads. The spin-transmission, or equivalently the spin-conductance $\sigma^s_{\phi \phi}$, is determined for a given direction $\phi$ at filling $\langle n \rangle$. 
    (b) The spin-resolved conductance as a function of angle $\phi$. In a trivial amorphous metal ($\triangle$) the conductance is isotropic and equally shared between the 2 spin species, whereas in the amorphous, metallic altermagnet ($\Square$ and $\pentagon$) the spin-resolved conductance is anisotropic. Total conductance is shown in grey. 
    (c) The spin splitter angle $\alpha$ quantifies the relative strength of spin conductance $G_\uparrow-G_\downarrow$ versus the total conductance $G_\uparrow+G_\downarrow$. At half-filling the system becomes insulating. Above half filling the sign of $\alpha$ is reversed because the upper two bands consist of $\ket{\uparrow y}$ and $\ket{\downarrow x}$ states, which have negative alter-magnetisation $m$. 
    }
    \label{fig:3}
\end{figure}%
{\it Spin-Conductance---}%
The central experimental feature of altermagnetism is non-zero spin transport. It reflects the presence of extended states which preferentially transport spin-up in the $x$-direction and spin-down in the $y$-direction or vice versa. We numerically evaluate the spin-resolved conductance $G_s$ through an amorphous sample, see \cref{fig:3}a, using the in-built Landauer--Büttiker algorithm of the \textsc{python} module \texttt{kwant} \cite{kwant}. The Landauer--Büttiker algorithm is based on an S-matrix approach, which relates the transmission probability of a metallic lead eigenstate with given momentum $k_x$, spin $s$, and orbital $\mu$ to the spin conductance $G_s$ \cite{landauer1957spatial,landauer1970electrical,buttiker1986fourterminal,buttiker1988symmetry}.

First, we focus on the total conductance $G_\uparrow + G_\downarrow$, shown in grey in \cref{fig:3}b. In accordance with the expectations for an amorphous system and our findings for the spectral function, the total conductance is close to isotropic, i.e., it is invariant under a global rotation $\phi$ of the system. We simulate global rotations of the system by rotating the internal degrees of freedom $T(\theta_{jk}) \rightarrow T(\theta_{jk}-\phi)$ which is in the thermodynamic limit equivalent to a rotation of the system. Slight deformations we dedicate to the finite-size shape anisotropy.

The spin-resolved conductance is strongly anisotropic. At low filling $(\Square)$ spin-up is preferentially transported in $x$-direction, and spin-down in the $y$-direction. At high filling $\pentagon$ these directions are swapped. The directional dependence of the spin-resolved conductance hence reveals the $d$-wave form factor of the spin splitting. 

Focussing on the spin conductance $G_\uparrow - G_\downarrow$ over the entire phase space, see \cref{fig:3}c, we observe that the spin conductance is only non-zero in the metallic altermagnetic phase, closely mirroring the phase diagram (\cref{fig:1}). In the low-filling limit the presence of a finite spin conductance is expected from a long-wavelength argument, reflecting the sharp features in the spectral function around the $\Gamma$-point. It is surprising that even at high filling---where the spectral function has only very limited interpretation and does not show any sharp features---the spin conductance is large. A quantitive analysis of the spin splitter angle $\alpha = 2 \arctan\left(\frac{G_\uparrow-G_\downarrow}{G_\uparrow+G_\downarrow}\right)$ shows that at both low and high filling there are regions of near-optimal splitting, $\alpha \sim 90^\circ$. Here, the system is almost completely insulating for one spin species in $x$-direction, while it is metallic for the other spin species. 

{\it Conclusion---}%
In summary, we have proposed a mechanism for realising an altermagnetic phase using only the spin and orbital degrees of freedom at each site. The altermagnetic symmetry group depends on the symmetry of the orbitals that are chosen, and not on any crystallographic symmetries. This allows us to construct altermagnetic phases on any non-crystalline system, as well as in crystalline systems where the altermagnetic symmetries can differ from the global rotational symmetry of the crystal. Additionally, the method constitutes a general procedure for constructing compensated collinear magnetic phases in amorphous materials. This provides a counterexample to the received wisdom that amorphous materials can only host ferromagnetic or glassy phases \cite{coey_amorphous_1978}.  
Our minimal model consists of itinerant electrons coupled via a ferromagnetic Kugel--Khomskii-like interaction, and has an amorphous altermagnetic ground state over a wide range of parameter space. The question of which materials, crystalline or amorphous, could realize this interaction microscopically or the instability in general remains open.

{\it Note added---}%
During preparation of this manuscript, we became aware of the recent related proposal of `atomic altermagnetism' \cite{jaeschke2025atomic}. We note that our instability presents an example of non-crystalline atomic altermagnetism and links it to spin-orbital ordering.

{\it Acknowledgments---}%
PD and AGG acknowledge fruitful discussions with Soren Bear, Frances Hellman, Joel Moore, Joe Orenstein, Nicola Spaldin, and Marc Vila. VL acknowledges support financial support from the ``Studienstiftung des deutschen Volkes''. AGG and PD acknowledge financial support from the European Research Council (ERC) Consolidator grant under grant agreement No.~101042707 (TOPOMORPH). JK~thanks for the hospitality of the Aspen Center for Physics, which is supported by National Science Foundation grant PHY-2210452. JK acknowledges support from the Deutsche Forschungsgemeinschaft (DFG, German Research Foundation) under Germany’s Excellence Strategy (EXC–2111–390814868 and ct.qmat EXC-2147-390858490), and DFG Grants No. KN1254/1-2, KN1254/2-1 TRR 360 – 492547816 and SFB 1143 (project-id 247310070), as well as the Munich Quantum Valley, which is supported by the Bavarian state government with funds from the Hightech Agenda Bayern Plus. JK further acknowledges support from the Imperial-TUM flagship partnership.

\bibliography{refs}

\appendix
\section{The Hopping Matrix}
\begin{figure}[t]
    \centering
    \includegraphics[]{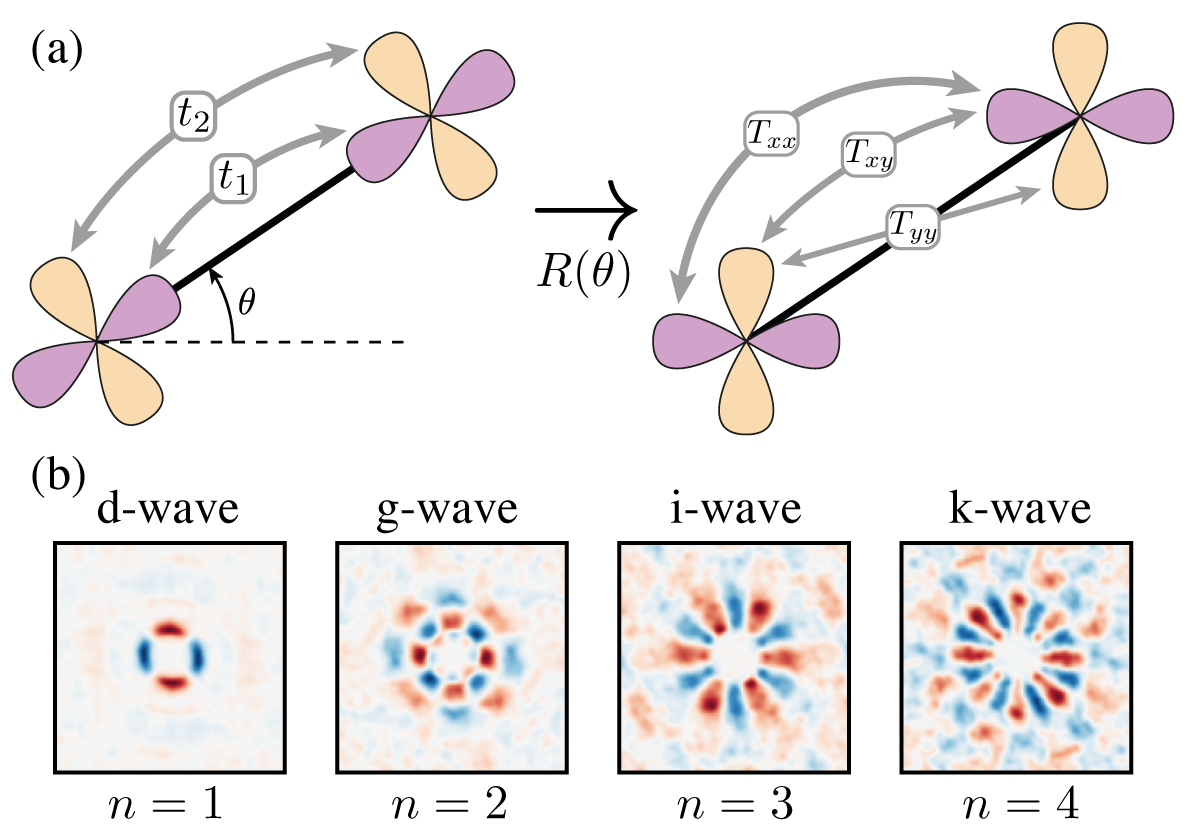}
    \caption{
    (a) The procedure for determining the hopping matrix $T(\theta)$. We start with a pair of orbitals rotated to align with the bond direction, assigning the hopping parameters $t_1$ and $t_2$. Rotating back to a canonical basis ($x$ and $y$ orbitals) we obtain the matrix elements of $T(\theta)$.
    (b) Spectral function for a series of higher-order altermagnetic phases, parametrised by $n$, where the altermagnetic symmetry of each phase is $C_{4n}\mathcal T$. Each phase was constructed and solved using the mean field prescription detailed in the main text. 
    } 
    \label{fig:t_matrix}
\end{figure}%

To produce a $d$-wave altermagnetic phase, we start with a pair of orbitals that transform into one another under a $\pi/2$-rotation about the $z$-axis. For example we could use $p_x$ and $p_y$ orbitals, or $d_{xz}$ and $d_{yz}$ orbitals, labelled by $x$ and $y$. 
Under a real-space rotation by $\theta$ around the $z$ axis, we may relate these two orbitals to a new basis, where the orbitals are parallel and perpendicular  to $\theta$,
\begin{align}
    \begin{pmatrix}
        c_{\parallel} \\
        c_{\perp}
    \end{pmatrix} = R(\theta) \begin{pmatrix}
        c_x \\
        c_y
    \end{pmatrix},
\end{align}
where $R(\theta)$ is the standard two-dimensional rotation matrix,
\begin{align}\label{eqn:rotation_matrix}
    R(\theta) = \begin{pmatrix}
        \cos\theta&  \sin \theta  \\
        -\sin \theta & \cos\theta
    \end{pmatrix}.
\end{align}
Thus, in order to calculate the $T(\theta)$ hopping matrix across a bond which makes an angle $\theta$ with the $x$-axis, we may start in a basis that is aligned with the bond, shown in \cref{fig:t_matrix}a. 
We assign a hopping $t_1$ between the two orbitals parallel with the bond, and a hopping $t_2$ to those that are perpendicular, arriving at the hopping term
\begin{align}
    h_{jk} = t_1 c_{j \parallel}^\dag c_{k \parallel}+
    t_2 c_{j \perp}^\dag c_{k \perp}+ h.c.,
\end{align}
where we have neglected to explicitly write the spin degree of freedom. Now we may rotate back to the global reference frame of $x$ and $y$ orbitals by applying the rotation matrix $R(\theta)$ on the orbital degrees of freedom on both on sites $j$ and $k$, arriving at a hopping of the form
\begin{align}
    h_{jk} = \begin{pmatrix}
        c_{jx}^\dag &
        c_{jy}^\dag
    \end{pmatrix}
    T(\theta)
    \begin{pmatrix}
        c_{kx} \\
        c_{ky}
    \end{pmatrix},
\end{align}
with
\begin{align}
    T(\theta) = 
    \begin{pmatrix}
        (t_1-t_2) \cos^2 \theta + t_2& 
        (t_1-t_2) \sin\theta \cos\theta \\ 
        (t_1-t_2) \sin\theta \cos\theta & 
        (t_1-t_2) \sin^2 \theta + t_2\end{pmatrix}.
\end{align}

\section{Higher Order Altermagnetism}

The extension to higher altermagnetic phases is straightforward. If we now consider using a pair of orbitals that transform into one another under a different fraction of $\pi$---for example $d_{xy}$ and $d_{x^2 - y^2}$ orbitals are mapped onto one another by a $\pi/4$ rotation about the $z$ axis---we see that the only change is that our effective rotation matrix, \cref{eqn:rotation_matrix}, acts on orbital space with a {\it multiple} $n \theta$ for $n\in \mathbb Z$. Thus, the only change to our Hamiltonian is that we now use hopping terms of the form
\begin{align}
    h_{jk} = \begin{pmatrix}
        c_{jx}^\dag &
        c_{jy}^\dag
    \end{pmatrix}
    T(n \theta)
    \begin{pmatrix}
        c_{kx} \\
        c_{ky}
    \end{pmatrix}.
\end{align}
For each value of $n$, the phase that arises is an altermagnet where the ground state respects a $ C_{4n}\mathcal T$ symmetry. 

Four examples of spectral functions corresponding to $n \in \left \{ 1,2,3,4 \right \}$ are shown in \cref{fig:t_matrix}b, corresponding to $C_4\mathcal T$, $C_8\mathcal T$, $C_{12}\mathcal T$ and $C_{16}\mathcal T$ altermagnetic phases. In each case we have constructed an altermagnetic Hamiltonian following the prescription of the main text, converging to an altermagnetic mean field ground state with $J=0.8$, filling $0.3$, $t_1 = 1$, $t_2 = 0.5$ on the same lattice geometry as that used in the main text, and determined the spectral function at the Fermi level. 
Note that half integer values of $n$ are forbidden, since they do not lead to collinear magnetic phases as discussed in \cite{smekjal_p_wave_magnets}, for example $C_6\mathcal T$ altermagnetism cannot be constructed.

\end{document}